\documentclass[prl,twocolumn,preprintnumbers,nofootinbib,superscriptaddress]{revtex4}
\usepackage{graphicx}
\usepackage{amssymb}

\usepackage{amsmath,amssymb,amsfonts,times,graphicx}
\usepackage[ps2pdf,bookmarks=true,colorlinks,linkcolor=red,urlcolor=blue,citecolor=blue]{hyperref}
\usepackage[usenames]{color}
\usepackage{dcolumn}
\usepackage{yfonts}

\def\({\left(}
\def\){\right)}
\newcommand{\be}{\begin{equation}}
\newcommand{\ee}{\end{equation}}
\def\aa{\mathfrak{a}}
\def\bb{\mathfrak{b}}
\def\hh{\mathfrak{h}}

\newcommand{\bwt}{\begin{widetext}}
\newcommand{\ewt}{\end{widetext}}

\begin{document}
\preprint{MIT -CTP/4291}
\title{Fractional Chern Insulators from
$\sqrt[n]{\text{Band Structure}}$}
\author{John McGreevy}
\affiliation{Department of Physics, Massachusetts Institute of Technology, Cambridge, MA 02139}
\author{Brian Swingle}
\affiliation{Department of Physics, Massachusetts Institute of Technology, Cambridge, MA 02139}\affiliation{Department of Physics, Harvard University, Cambridge, MA 02138}
\author{Ky-Anh Tran}
\affiliation{Department of Physics, Massachusetts Institute of Technology, Cambridge, MA 02139}

\begin{abstract}
We provide a parton construction of wavefunctions and effective field theories for fractional Chern insulators.  We also analyze a strong coupling expansion in lattice gauge theory that enables us to reliably map the parton gauge theory onto the microsopic Hamiltonian.  We show that this strong coupling expansion is useful because of a special hierarchy of energy scales in fractional quantum Hall physics.  Our procedure is illustrated using the Hofstadter model and then applied to bosons at $1/2$ filling and fermions at $1/3$ filling in a checkerboard lattice model recently studied numerically.  Because our construction provides a more or less unique mapping from microscopic model to effective parton description, we obtain wavefunctions in the same phase as the observed fractional Chern insulators without tuning any continuous parameters.
\end{abstract}
\maketitle

\section{Introduction}
The discovery of the fractional quantum Hall effect in layered semiconductor devices in high magnetic field
\cite{FQHE}
opened our eyes to a rich
world of truly quantum mechanical phases of matter that exist at zero temperature. Quantum entanglement plays a crucial role
in defining and understanding topological phases of matter like the fractional quantum Hall liquids.  Instead of long range order and symmetry breaking, we should study the pattern of long range entanglement in such phases
\cite{WenNiu, Xiao-GangWenbook}).  One manifestation of the necessity of quantum entanglement is the inability of product or mean-field like wavefunctions to capture, even qualitatively, the physics of such a phase. Their potential ability to function as quantum computers \cite{topreview}, as well as their natural robustness to local decoherence, are also consequences of the presence of long range entanglement in the fractional quantum Hall fluids.

Despite the beauty and ruggedness of these topological liquids, there are still many practical challenges as we attempt to observe non-Abelian particles in nature and construct the first scalable quantum computer. The zero-temperature robustness of these topological fluids
is eventually lost at finite temperature, and since all experiments are carried out at finite temperature, we are forced into practical questions.  Exactly how big is the gap to excitations?  Just how easy is it to implement the non-local operations that detect non-Abelian excitations or perform useful quantum computations?  Moreover, given the conceptual and practical importance of topological phases of matter, it is very interesting to ask where else in nature such topological phases may be lurking.

Motivated by these questions and others, there has been an explosion of interest in new models that mimic the physics of the quantum Hall effect.  In the Hall effect, the basic starting point is Landau levels, perfectly flat bands that exist in a uniform magnetic field.   When Landau levels are totally filled we find the integer quantum Hall effect, and when Landau levels are partially filled, there exists a large degeneracy in the non-interacting limit.  The inclusion of interactions resolves the degeneracy and produces an incompressible topological fluid \cite{laughlin}, a fractional quantum Hall fluid.  The physics of Landau levels may be conjured anew on the lattice by studying tight binding models that break time reversal symmetry \cite{Haldanehoneycomb, HatsugaiKohmoto}.  Such models can be tuned to give very flat bands which simultaneously possess a non-trivial Chern number. Filled bands with non-trivial Chern number are in the same universality class as integer quantum Hall states, and the phenomenology is largely transferable.  Thus it is reasonable to guess that if we were able to partially fill a flat Chern band and expose the electrons to interactions of a magnitude much greater than the bandwidth, then the physics of the fractional quantum Hall effect should also be visible (see Ref.~\cite{Physics.4.46} for a discussion of this point).  This beautiful idea, which is a striking example of universality in physics, has recently been numerically confirmed in a flurry of activity \cite{Neupert, Sheng11, evelyn, hardcore-bosons, bernevig}.  (Previous work, motivated by 
the possibility of cold atoms realizations of FQHE includes 
\cite{demler,
moeller_composite_2009}.)

From the perspective of the questions raised above, these fractionalized Chern insulators are interesting in that they provide a new realization of quantum Hall physics with the potential for much higher energy gaps and the possibility of new methods for the detection anyons.  Particularly interesting is the idea of realizing new high temperature non-Abelian topological liquids, on which we comment later. As an important early step towards the observation and manipulation of non-Abelian anyons in fractional Chern insulators, we must establish a basic theoretical framework in which to understand numerical and experimental results.  We have in mind the powerful formalism in fractional quantum Hall physics known as the slave-particle or parton approach that provides wavefunctions and low energy effective theories for fractional Hall liquids ({\it e.g.}~\cite{Xiao-GangWenbook,Fradkin:1991nr,PhysRevLett.66.802,PhysRevB.60.8827}).  In practice, the parton approach to this problem represents a rationalization for and a massive generalization of Laughlin's original wavefunction \cite{laughlin_wf} for the $\nu=1/3$ plateau.  We are interested in transferring this technology to the new setting of fractional Chern insulators.  An interesting recent attempt to make this transition and provide wavefunctions for fractional Chern insulators can be found in Ref.~\cite{XLQwavefunction}.  Parton wavefunctions similar to those we provide have been used in variational studies in Ref.~\cite{2011arXiv1102.2406M}.  Ref.~\cite{Shankar-composite-fermions} also uses technology from quantum Hall by embedding the Chern band problem into a larger Landau level.  
Ref.~\cite{sondhi} extends a $W_\infty$ structure of quantum Hall states to Chern bands.

In this paper we achieve this aim by formulating a parton description of the recently observed Abelian fractional Chern insulators.  This parton description provides model wavefunctions and low energy effective theories that demonstrate the universality of fractional quantum Hall physics.  The core of the parton formalism is a mapping of the original electronic system to an effective description in terms of fractionally charged ``partons" coupled to emergent gauge fields.  Analysis of a parton model always requires analysis of a gauge theory, and in the case of fractional Chern insulators, we deal with lattice gauge theory.  In addition to the basic story involving wavefunctions, we formulate and carry out a strong coupling expansion of certain lattice gauge theories.  This strong coupling analysis\footnote
{Deriving the dynamics of hadronic objects using a strong coupling expansion 
in lattice gauge theory dates back at least to \cite{PhysRevD.22.490} 
and is used in the same spirit in {\it e.g.}~\cite{MotrunichFisher}.} 
yields two immediate benefits: new insights into the dynamics of some strongly coupled gauge theories, 
and new clues leading to fractional Chern insulators, 
including novel non-Abelian states. In fact, because the strong coupling expansion provides an almost unique mapping from the microscopic Hamiltonian to the lattice gauge theory, our construction can be regarded as a zero parameter model of the original microscopic physics in terms of the gauge degrees of freedom.  Because the gauge theory dynamics can be reliably analyzed (thanks to the Chern-Simons term), our construction provides strong evidence that the system enters a fractional Chern insulating phase.

As we explain in more detail below, the main conceptual obstacle to the construction of a parton theory for fractional Chern insulators is the question of the parton band structure.  In the case of Landau levels, the degeneracy of each Landau level naturally depends on the charge of the particle moving in the magnetic field.  Consider the case of $\nu = 1/3$.  Breaking the electron into three ``colors" of charge $1/3$ partons leads to a partonic Landau level with reduced degeneracy, and since the number of partons of each color is the same as the number of electrons, the partons are immediately able to completely fill their Landau level.  This picture leads to an easy accounting of the Hall conductivity, as we have three colors of partons each contributing a Hall conductivity of $\frac{(e/3)^2}{h}$ for a total Hall conductivity of $\frac{1}{3} \frac{e^2}{h}$.  Vaezi \cite{Vaezi} has noticed that the story for fractional Chern insulators is not as simple, since it is less straightforward to reduce the size of the partonic bands, but the state Vaezi constructs is not obviously gauge invariant and appears to give zero after projection.  We address this issue in our construction by expanding the size of the unit cell seen by the partons in a way that remains invisible to the microscopic electrons.  This discrete set of choices is the only freedom in the parton band structure that is needed to open a gap in the single particle spectrum.

The remainder of the paper is organized as follows.  First, we describe our construction in the context of a simple lattice model of the fractional quantum Hall effect, namely the partially-filled Hofstadter model.  In the same section we discuss a strong coupling expansion for the parton gauge theory that, owing to the special physics of the fractional quantum Hall effect, leads to an unusually complete physical justification for the parton approach.  Second, we apply our construction to a case that makes contact with recent numerical calculations, fractionally filled fermions and fractionally filled bosons moving in a checkerboard model.  Finally, we summarize our results and indicate some exciting directions we are currently exploring.

Note: while this paper was in preparation, we learned of a paper by Lu and Ran \cite{LuRan} which also develops a parton construction for fractional Chern insulators.  The overlap between their work and ours is significant; our discussion of the strong coupling expansion is a substantial difference.  Their work is a general discussion of a variety of parton wavefunctions including time reversal symmetric insulators, while we focus on the story of the strong coupling expansion and the case of fractional Chern insulators.

\section{Hofstadter model and strong coupling expansion}
We introduce our construction in the simple context of the Hofstadter model of particles hopping in a uniform magnetic field on a lattice \cite{PhysRevB.14.2239}.  The Hofstadter model is defined by taking a square lattice and placing $2 \pi / N$ flux through each plaquette of the lattice.  For a fixed $N$, the Hofstadter model has a unit cell consisting of $N$ sites and thus a band structure consisting of $N$ bands.  As $N\rightarrow \infty$, the lower bands of the Hofstadter model reproduce the continuum physics of Landau levels.  In terms of universal physics, the lowest band of the Hofstadter model has Chern number one and hence reproduces all the low energy physics of the $\nu =1 $ integer quantum Hall state even at finite $N$.  Similarly, a fractionally filled lowest Hofstadter band in the presence of interactions is a natural lattice regularizaton of a fractional quantum Hall fluid.  We now consider the case of electrons moving in the lowest Hofstadter band at filling $1/3$.

Consider the amplitude for an electron to hop around one plaquette of the square lattice.  This amplitude has a phase given by $e^{2 \pi i/N}$ which we interpret as an Aharanov-Bohm phase for an electron of charge $e$ moving in a field of magnitude $B = \frac{2 \pi}{N e a^2}$ where $a$ is the lattice spacing.  Although the models discussed in \cite{Neupert, Sheng11, evelyn, hardcore-bosons, bernevig} generally speak of quantum Hall type systems without magnetic fields, we see that we can always reinterpret phases among the hopping parameters in terms of strong lattice scale magnetic fields.  This interpretation is useful because it suggests a way to obtain a sensible parton band structure.  If the electron fractionalizes into charge $e/3$ partons then these partons should naturally experience a phase of $e^{2 \pi i/(3 N)}$ when moving around a plaquette because of their reduced charge.  Thus at the mean field level, before including the gauge fluctuations that glue the partons back together into electrons, the partons move in a modified Hofstadter bandstructure with $N$ replaced by $3N$.  This means the partons move in an enlarged unit cell relative to the electrons, and with one parton of each color per electron, each parton color can fill an entire band, leading to a gapped mean field ground state.

Despite the apparent enlargement of the unit cell, the physical electron does not experience an enlarged unit cell since it encircles $2 \pi$ flux after moving around only $N$ plaquettes.  Equivalently, motion of the electron requires coordinated motion of the partons, and counting the phase of $2 \pi / 3$ per color accumulated by a parton around $N$ plaquettes, we find a total phase $2 \pi$ after summing over the three colors.  The analogy to the usual continuum story
\cite{Xiao-GangWenbook,PhysRevLett.66.802,PhysRevB.60.8827}
is, we hope, now clear.  In particular, the electron wavefunction is obtained by taking the mean field wavefunction of the partons, three copies of the lowest $3N$ Hofstadter band, and projecting onto the color-neutral state.  As we take $N \rightarrow \infty$ this procedure recovers Laughlin's wavefunction for $\nu = 1/3$.

\begin{figure}[h] \begin{center}
\includegraphics[width=.48\textwidth]{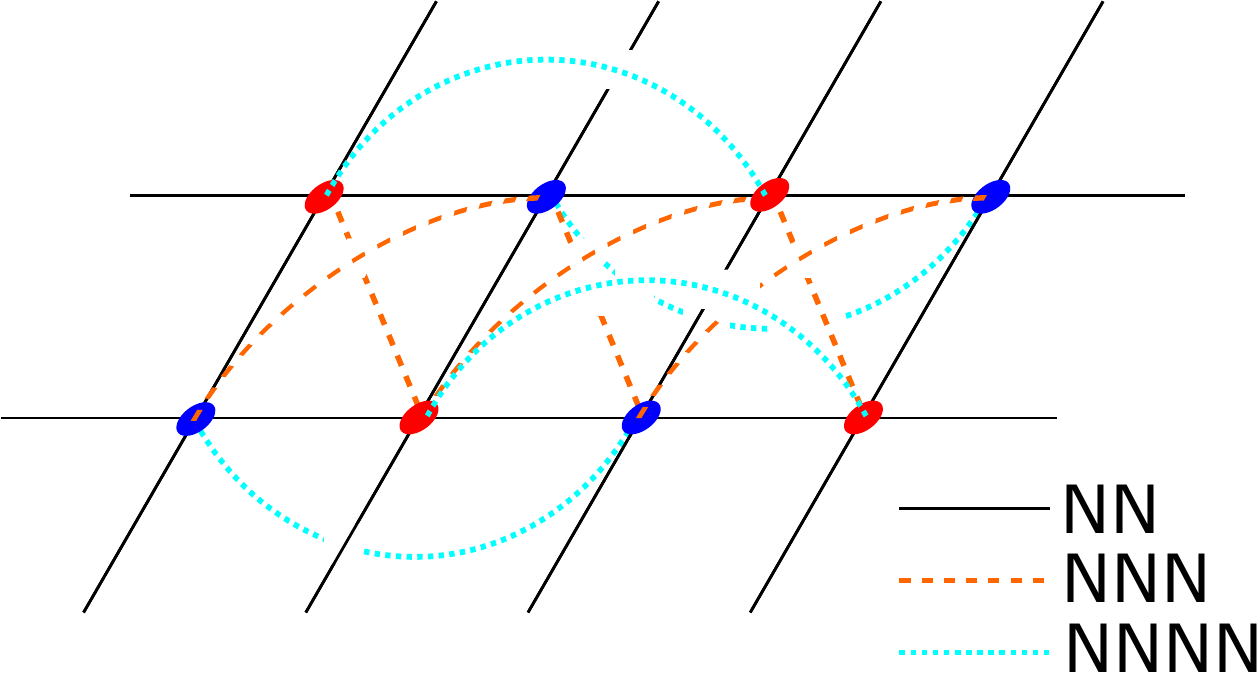}
\end{center}
\caption{
\label{fig:checkerboard_3d}
A rendition of the checkerboard model showing the NNNN hoppings interpreted as ``bridges" out of the plane connecting more distant sites.  The red and blue dots are the two sublattices.  Our lattice gauge theory version of this model associates an $SU(3)$ link variable with every elementary hopping path shown here.  Included in the sum over loops are loops running along these out of plane bridges representing the NNNN hoppings.}
\end{figure}

The above procedure may seem somewhat ad hoc, but there is a more systematic way to relate the parton band structure to the electron band structures from which it emerges.  First, we formalize the parton construction by writing the electron operator $c_r$ at site $r$ in terms of three partons $f_{r \alpha}$ carrying charge $e/3$ as $c_r = f_{r 1} f_{r 2} f_{r 3} = \frac{1}{3!} \epsilon_{\alpha \beta \gamma} f_{r \alpha} f_{r \beta} f_{r \gamma} $ as in Ref.~\cite{PhysRevLett.66.802}.  This expression is manifestly symmetric under $SU(3)$ transformations of the $f_{r \alpha}$, and since we can make such transformations at each site $r$, we have an $SU(3)$ gauge structure.  The group $SU(3)$ is the so-called high energy gauge group.  The terminology is necessary because the mean field parton Hamiltonian may break some of the $SU(3)$ symmetry, so the true low energy gauge group can be distinct from $SU(3)$.  Next, we write a Hamiltonian for the parton theory including the $SU(3)$ gauge fluctuations.  For every term in the parton hopping Hamiltonian we associate an $SU(3)$ link variable $V$.  This particular choice is very convenient for performing a strong coupling expansion, and can be visualized as allowing the partons to hop longer distances through lattice scale ``bridges" or ``wormholes" as shown in Fig. \ref{fig:checkerboard_3d}.  Our notation is as follows: the lattice is defined by a graph $(\mathcal{V},\mathcal{E})$, fermions $f_r$ live on the sites $r \in \mathcal{V}$ and gauge fields $V_{rr'}$ live on links $rr' \in \mathcal{E}$.  There is a link in $\mathcal{E}$ for every elementary hopping term in the parton Hamiltonian.  The set of elementary loops on the lattice we denote $\mathcal{L}$.  The (chromo)electric field conjugate to $V_{rr'}$ is $E^A_{rr'}$ where $A$ is an adjoint index for $SU(3)$.  With these conventions, our Hamiltonian is
\begin{eqnarray} 
H = - \sum_{rr' \in \mathcal{E}} t^f_{r r'} f^+_{r'} V_{r r'} f_r + \mbox{h.c.} + h \sum_{rr' \in \mathcal{E}} E^2_{rr'} \nonumber \\ - K \sum_{\ell \in \mathcal{L}} \mbox{tr}\left(\prod_{rr' \in \ell} V_{rr'}\right) +\mbox{h.c.}
\end{eqnarray}
The first term is a parton hopping term, the second electric term favors small electric fields, and the third magnetic term favors smooth gauge configurations.  $E^2_{r r'}$ is the quadratic Casimir of $SU(3)$ on the link $r r'$, so the eigenstates of the electric field term are labelled by irreducible representions of $SU(3)$.   In this basis, the magnetic field term functions like a raising and lowering operator: it adds an electric field line in the fundamental along the path $\ell$.  

Note that the physical Hilbert space of the gauge theory may be formally larger than the electron Hilbert space due to the presence of arbitrarily highly excited gauge fields states or ``glueballs".  Thus the gauge theory technically describes electrons coupled to some high energy tower of bosonic states, however, these bosonic degrees of freedom are irrelevant for the low energy physics.  We thus proceed to study the gauge system at strong coupling as a model of the low energy physics of electrons.

\begin{figure}[h] \begin{center}
\includegraphics[width=.48\textwidth]{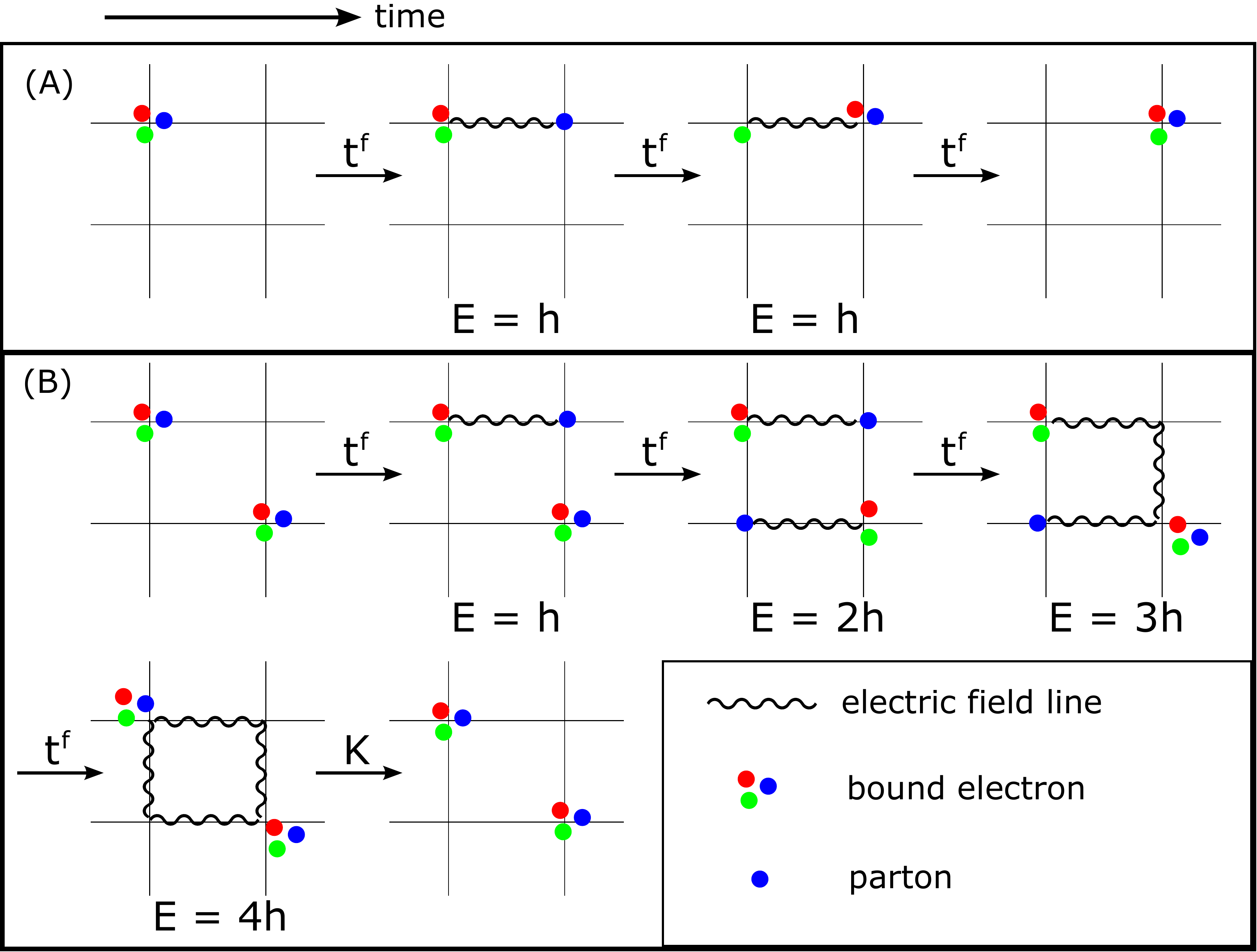}
\end{center}
\caption{
\label{fig:strong_coupling}
The two most important processes in the strong coupling expansion.  Process (A) generates an electron kinetic term of order $(t^f)^3/h^2$.  Process (B) generates repulsive electron interactions of order $(t^f)^4 K /h^4$.}
\end{figure}

Let us analyze this lattice gauge theory in a strong coupling expansion $h \gg K, t^f_{r r'}$.  We will need only the lowest order terms for our purposes here, but the expansion can be carried out systematically to higher orders where it is important to remove disconnected terms \cite{RevModPhys.39.771}.  The excitations that survive in the $h\rightarrow \infty$ limit are the colorless fermionic ``baryons" generated by the operators $c_r = f_{r 1} f_{r 2} f_{r 3}$, that is, the original electrons.  Away from $h = \infty$, these electrons can hop via virtual fluctuations of their constituent partons.  Consider a particular parton hopping amplitude $t^f$ and suppose we begin with an electron on site $r$.  The parton hopping Hamiltonian moves one parton from site $r$ to site $r'$ with amplitude $t^f$, but it also creates an electric field line connecting $r$ to $r'$.  Thus this intermediate state costs an energy of roughly $h$.  We now proceed to hop a second parton with the resulting intermediate state still containing electric fields.  Finally, we hop the third parton, remove all electric fields, and return to a colorless state at $r'$.  This process is illustrated in panel $(A)$ of Fig. \ref{fig:strong_coupling}.  In effect, the electron moved from $r$ to $r'$ with an amplitude given by $t^c \sim (t^f)^3/h^2$.  This argument applies to all the parton hopping amplitudes, so we have for every parton hopping amplitude a corresponding electronic hopping amplitude given by $(t^f)^3/h^2$ (with the conventional overall minus sign).  Furthermore, in our scheme the prefactor is the same in all cases (only one link ever has an electric field excited).  Of course, there are other contributions to the electron amplitude from more complicated processes, but these are further suppressed by powers of $t^f/h$.  Note that in this analysis we
focus on the interactions between two baryons and ignore
possible many-body effects from the background density of baryons;
this should be justified since such effects will require more complicated parton exchange
patterns and will be further suppressed.

If we express all electron hoppings in terms of a particular matrix element $t^c_0$, say a nearest neighbor hopping, then we have the remarkable result that $\frac{t^c_{r r'}}{t^c_0} = \left(\frac{t^f_{r r'}}{t^f_0} \right)^3$.  In other words, the electron hoppings measured relative to a reference hopping are literally the cubes of the parton hoppings measured relative to the equivalent parton reference hopping.  Hence the parton bandstructure represents, in a very precise sense, the ``cube root" of the electron band structure.  It is important that the mapping from electrons to partons is not quite unique, we always have discrete phase choice not set by the electron band structure.  In the case of the Hofstadter model, this prescription is very simple since there is only one kind of hopping term, and so, up to the overall energy scale, we reproduce our simple guess that the parton Hamiltonian should have a third of the phase of the electron Hamiltonian through every plaquette.

So far we have only mentioned electron kinetic terms.  Electron interactions are also generated.  A simple process that generates such interactions is parton exchange between two neighboring electrons as shown in panel $(B)$ of Fig. \ref{fig:strong_coupling}.  This process involves not only parton hopping but also the magnetic term to remove the final closed field line, and so it has an amplitude $U^c \sim K (t^f)^4/h^4$ and is repulsive (an extra minus sign comes the fermion exchange).  In the strong coupling limit electron interactions are parametrically smaller than electron kinetic terms.  Interactions are also short ranged since parton exchange over a distance $x$ is supressed by factors of $(t^f/h)^x$.  Note also that the magnetic term in the gauge Hamiltonian generates ring-exchange terms for the electrons \cite{MotrunichFisher}, where electrons hop around loops in a coordinated fashion.  These terms are of order $K (t^f/h)^6$ and can be parametrically smaller than $t^c$ and $U^c$.  Now the usual story at this point would be simple: the gauge field confines and we can integrate it out to produce a weakly coupled electronic Hamiltonian which can be analyzed in the usual way.  We emphasize that confinement is a very common situation in gauge theory, and deconfinement typically requires special circumstances: high density, many flavors, broken time reversal, etc.  Since strongly correlated electron systems typically need electron interactions of the same order as electron kinetic terms, such a phase corresponds to the medium coupling region in the gauge theory $h \sim t^f \sim K$.  The usual hope is that once we make $K$ and $t^f$ large enough, the gauge theory will flow in the infrared to a deconfined phase, but we emphasize that because the strong coupling expansion breaks down at intermediate coupling, we cannot use it to reliably predict electronic models that realize such deconfined phases.  Of course, it remains invaluable as a source of intuition.

However, the case of fractional Chern insulators is special.  This is because although the bare electron hoppings are much larger at large $h$ than the electron interactions, the bandwidth of the nearly flat electronic band can be tuned to be quite small.  We use the Hofstadter model as an example.  Let the electronic gap be $\Delta^c \sim t^c$, let the bandwidth of interest be $w^c$, and suppose it is tuned to be some small fraction $F$ of the electron hopping
(in the Hofstadter model, this can be accomplished by making $N$ large;
more generally it is the goal of much bandstructure engineering \cite{PhysRevB.78.125104,Sun10}).  Then we want to implement the hierarchy of scales $\Delta^c \gg U^c \gg w^c$.
This can in fact be achieved within a region of parameter space where the strong coupling expansion is reliable, so long as $F \ll 1$.  All we need is $h \gg \sqrt{t^f K} $ and $h \ll \sqrt{\frac{t^f K}{F}}$.  Of course, unless the electronic band is perfectly flat, we will eventually violate the hierarchy if we take $h$ to infinity or $K$ to zero while fixing $t^f$.

Based on the model systems studied in Refs. \cite{Neupert, Sheng11, evelyn, hardcore-bosons, bernevig}, a reasonable target Hamiltonian for realizing a fractional Chern insulator should have the following features:
\begin{enumerate}
\item{} Large electronic band gap: $\Delta^c \gg w^c, U^c$,
\item{} Strong electron interactions: $U^c \gg w^c$,
\item{} Small ring exchange terms.
\end{enumerate}
Regarding criterion 3, it is usually believed that ring exchange terms are helpful in the fight against confinement and we have no reason to suspect otherwise here. Nevertheless, the most conservative path is to limit ourselves as much as possible to the minimal ingredients present in numerical studies that realize fractional Chern insulators.\footnote{We should point out that in the $\nu = 1/2$ boson case 
described below, the ring exchange term is of the same order as the interactions.}  Of course, ring exchange terms will be generated under renormalization regardless of whether we include them in the microscopic Hamiltonian.  We can realize all three criteria within the strong coupling expansion by establishing the hierarchy $\sqrt{(t^f K)/F} \gg h \gg K \gtrsim t^f$.  This guarantees the electronic hierarchy $\Delta^c \gg U^c \gg w^c$ plus small ring exchange terms.

We are now in a privileged position.  We have a precise and controlled map between the gauge theory and the microscopic electron theory via the strong coupling expansion.  We also know the fate of the electron theory: it forms an incompressible phase with the same universal physics as the $\nu = 1/3$ fractional Hall fluid.  Thus we also know the fate of the gauge theory: it must deconfine.  To be fair, it is possible that the particular interactions generated somehow always favor a crystalline state, but we regard this as very unlikely.  Thus deconfinement must occur even if $t^f$ and $K$ are relatively small.  As we said above, $h$ cannot be infinite and $K$ cannot be zero, but our partonic ``molecules" can be relatively tightly bound in isolation and still deconfine when put together into a liquid.  We note that there exist duality arguments that suggest that a lattice Chern-Simons term pushes the confinement transition to $h = \infty$ \cite{Rey1991897}.  A hint that this is sensible comes from the continuum limit where the Chern-Simons term (generated by the partons \cite{Redlich1, Redlich2}) leads to a mass gap
\cite{Deser:1981wh, Deser:1982vy}
no matter the size of the continuum gauge coupling \cite{Grignani:1996ft,Fradkin:1994tt,PhysRevLett.66.276}.  In the compact Abelian gauge theory, Polyakov's argument for confinement 
\cite{Polyakov:1975rs} breaks down because the Chern-Simons term 
attaches fermion charge to the instantons;
they therefore only contribute to fermion correlation functions, rather than to the free energy.

The same physics is evident on the lattice, where the collective motion of the partons responsible for the Hall conductivity \cite{TKNN} is not confined, even at large $h$, precisely because it is adiabatic.  Furthermore, because the Hall conductivity can be expressed without reference to excitations above the partonic band gap \cite{TKNN}, it is in some sense independent of $t^f$ and always contributes an order one effect (like the continuum Chern-Simons term).  Of course, once we move away from the low frequency response, we will eventually encounter the details of the parton band structure.  
We note that while the lattice is essential for dealing correctly with $U(1)$ gauge theory, it can effectively be dispensed with when considering non-Abelian groups.  While the naive continuum limit of $U(1)$ gauge theory fails to incorporate the instanton effects that lead to confinement and hence looks free, non-Abelian gauge theories in the continuum can already accommodate confinement.  Thus we expect that the continuum model with Chern-Simons term correctly captures the physics even at strong gauge coupling.

As further evidence of the similarity between the lattice and the continuum, we emphasize that the deconfined phase of the lattice gauge theory does not have light propagating pure gauge excitations.  Exactly as in the continuum, the strong gauge coupling leads to very heavy pure gauge states, namely the high energy ``glueballs" we mentioned earlier in connection with the physical Hilbert space.  These excitations occur at energy scale $h$.  What we must have in the low energy physics are renormalized partons that do not experience confining electric fields, as well as certain zero modes of the gauge field.  In conjunction with the TKNN invariant, these gauge field zero modes lead to ground state degeneracy on topologically non-trivial surfaces.  Finally, regarding the partons, it is useful to appeal to the bag model of confinement (see {\it e.g.}~\cite{Johnson:1975zp}).  Imagine we start with a dilute gas of electrons where each electron may be modeled as a ``bag" of the deconfined phase.  The partons locally experience their band structure plus the confining potential of the bag, hence the partons propagate along the edge of the bag (a manifestation of the edge states in a finite size Chern insulator).  When the electrons reach a density of order one, the different bags strongly overlap and partons may tunnel from one bag to another.  Analagous (but not identical) to the percolation transition between quantization plateaus in the integer quantum Hall effect, the partons are ultimately able to connect up and form a gapped collective state.  We may picture the partons moving along the edges of the bags to screen out any charge sources. Of course, such a state does not screen in the same way as a metal, but a type of screening is certainly present.  The simplest demonstration of this fact comes from the abelian equations of motion, where all static fields are short ranged.

We offer one more argument for the deconfined nature of the lattice gauge theory.  We can consider a microscopic electron model that possesses a completely flat Chern band at the cost of introducing long range hoppings.  As we argue in Appendix A, these hoppings may be chosen to decay super-polynomially fast or nearly exponentially fast (although perhaps not exponentially fast).  Now in such a model, the condition $U^c \gg w^c$ on the electron side is vacuous, and we appear to be able to take $h$ arbitrarily large on the gauge theory side.  Yet surely such a model still realizes the fractional Chern insulator phase (at least if the partons remain gapped), since it closely mimics the physics of the usual continuum quantum fractional Hall effect.  Thus we must have that the gauge theory deconfines even if $h$ is large.  We are led to conclude that fermions in a filled Chern band and the Chern-Simons term they generate have a profound impact on the dynamics of the gauge fields they are coupled to.

Note that infinite range hoppings do not require a proliferation of gauge degrees of freedom.  This is because instead of introducing new gauge fields for each hopping, we may simply stretch a Wilson line between the hopping sites.  This Wilson line only reduces the electron hopping amplitude by a factor of the distance, a modification that can easily be absorbed into a redefinition of the parton hoppings without changing their super-polynomial decay.

Having dealt with the issue of confinement, we turn to the low energy physics.  Although it is not our purpose here to reanalyze the low energy theory (see Refs. \cite{Xiao-GangWenbook,PhysRevB.81.155302}), we do want to make a few comments.  For the Hofstadter model, the low energy limit of the lattice gauge theory consists of a non-Abelian $SU(3)$ gauge field with a Chern-Simons $k \int \(A d A + \frac{2}{3} A^3 \)$ term at level $k = C_{\mbox{parton}} = 1$.  By level-rank duality, this is equivalent 
in the bulk to a $U(1)$ Chern-Simons theory at level $k=3$ which possesses three degenerate ground states on a torus.  The low energy theory also contains Chern-Simons term for the background gauge electromagnetic gauge field that encodes the Hall conductivity.  The theory contains gapped fermionic matter, the partons, coupled to the $SU(3)$ gauge field, and the usual anyonic statistics follow from a more detailed analysis.

To summarize this section, we have introduced a strong coupling expansion that justifies our procedure for taking the ``$n$th root of band structure". Furthermore, we showed that the unique physics of the fractional quantum Hall effect permits an unusually complete understanding of the connection between the microscopic electron model and its partonic gauge theory description. Since the gauge theory can be reliably analyzed, our construction provides a strong argument that the microscopic electron model enters a fractional Chern insulating phase.  As an application of this technology, we can with reasonable confidence propose microscopic Hamiltonians that may realize non-Abelian fractional Chern insulators.  We now turn to the case where, for definiteness, most of the attention has been focused --  the partially filled checkerboard model -- and describe in detail the partonic band structures and wavefunctions generated by our approach.

\section{Checkerboard model parton construction}

The desiderata for the $n$th root operation
to provide a good fractional Chern insulator mean-field state are:
\begin{enumerate}
\item{} Gauge-invariant quantities should be invariant under the
translation invariance of the original lattice model, with the original unit cell.
This ensures that the resulting physical wavefunction is translation invariant.

\item{} There should be a bandgap between the filled and empty bands.
The filling fraction for each parton color is the same as for the original particles.

\item{}\label{item:chern} The total Chern number of the filled bands should be nonzero.

\end{enumerate}

The final criterion \eqref{item:chern} is crucial in order
to prevent the gauge theory from entering a confining phase.

In the previous section we described a strong coupling perturbation theory
calculation that narrows our search:
define a new lattice with the same connectivity structure as the original,
and define the hopping amplitude on a given link to be the $n$th root
of the original amplitude.  This leaves a ${\bf Z}_n$ phase ambiguity on each link;
choosing these phases to all be $+1$ inevitably leads to partially filled parton bands;
choosing these phases to break lattice translation invariance
to an order-$n$ subgroup produces a new bandstructure
which can meet the above criteria.  We emphasize again that the hierarchy of scales considered in
\cite{Sun10, Neupert, Sheng11, evelyn, hardcore-bosons, bernevig} is precisely what is needed to have a reliable strong coupling expansion.  Thus our $n$th root procedure is, modulo the important discrete freedom described above, a unique and precise mapping from the microscopic model to the gauge theory.  For the moment we proceed with a general description keeping $n$ arbitrary; later we will specialize to bosons ($n=2$) and fermions ($n=3$).

Consider a tight-binding model defined by the following hamiltonian 
\be H = - \sum_{rr' \in \mathcal{E}} t^c_{rr'} c^\dagger_{r} c_{r'} + {\text{h.c.}} \ee
Here $c^\dagger, c$ denote creation and annihilation operators
for electrons or hardcore bosons; the sites $r$ are drawn from the set $\mathcal{V}$ which is a Bravais lattice plus basis.
Our mean field ansatz for the $n$th root of this lattice is
\be H_{\mbox{mf}} = - \sum_{\alpha} \sum_{rr'\in \mathcal{E}} \( t^c_{rr'} \)^{1/n} \omega_{r r'} f^\dagger_{r \alpha} f_{r' \alpha} + {\text{h.c.}} \ee
$f^\dagger_{\alpha}, f_{\alpha}$ create and annihilate $n$ colors (labelled by $\alpha$) of fermionic partons,
from which the original electrons or hardcore bosons are constructed as
baryons, $ c_r = {1\over n!} \epsilon_{\alpha_1...\alpha_n} f_{r \alpha_1} ... f_{r \alpha_n}$.
At this point, we have not yet included the gauge fluctuations, so the color label in the mean field Hamiltonian above should be treated like a flavor or band index.  We assume that the mean field Hamiltonian respects the full $SU(n)$ symmetry.  $\omega_{r r'}$ are $n$th roots of unity
expressing the ambiguity in the $n$th root operation; they can be regarded as variational parameters. Not all of them can be eliminated by rephasing the parton operators $ f_{r \alpha} \to \omega_r f_{r \alpha}$ ($\omega_r$ independent of $\alpha$).
It is important to note that although the graph is the same as that of the original model, the unit cell of the parton mean field Hamiltonian may be expanded relative to the original due to the phases $\omega_{r r'}$.

For definiteness, we focus on the checkerboard model
studied by Sun {\it et al} \cite{Sun10,PhysRevLett.103.046811}.
%
This is a tightbinding model with two sites per unit cell
and hopping amplitudes out to the next-next-nearest-neighbor (4N).
Electrons in this bandstructure at
${1\over 3}$ filling
(that is, one electron for every 3 unit cells, or every 6 sites)
with nearest-neighbor repulsive interactions,
exhibit an insulating state
\cite{Neupert, Sheng11, bernevig}
as do repulsive hardcore bosons at ${1\over 2}$ filling
\cite{hardcore-bosons}.
The electron model has been studied
for system sizes
up to $12$ electrons on $36$ unit cells \cite{bernevig},
and at various values of the hopping amplitudes
including $t''=0$ (``$\pi$-flux" model of \cite{Neupert}).

These works find evidence for $1/\nu$ degenerate groundstates
on the torus (in each case, $\nu$ is the number of particles per unit cell),
which flow into each other under threading of flux.  This is precisely the finite size signature of a fractional quantum Hall liquid, however, we note that an electronic charge density wave which breaks lattice translation symmetry also shows the same qualitative physics.  Indeed, in the limit of a thin torus, the quantum Hall fluid actually takes the form of a one dimensional CDW \cite{SeidelLee1, SeidelLee2}.  However, on a torus of aspect ratio near one, the mixing between the three ground states are very different in the FCI and CDW cases; furthermore, the entanglement spectra will also be very different.  Following the numerical calculations and our analytic results above, we assume the state is a FCI and go on to provide candidate wavefunctions for each of these states.

\begin{figure}[h] \begin{center}
\includegraphics[height=100pt]{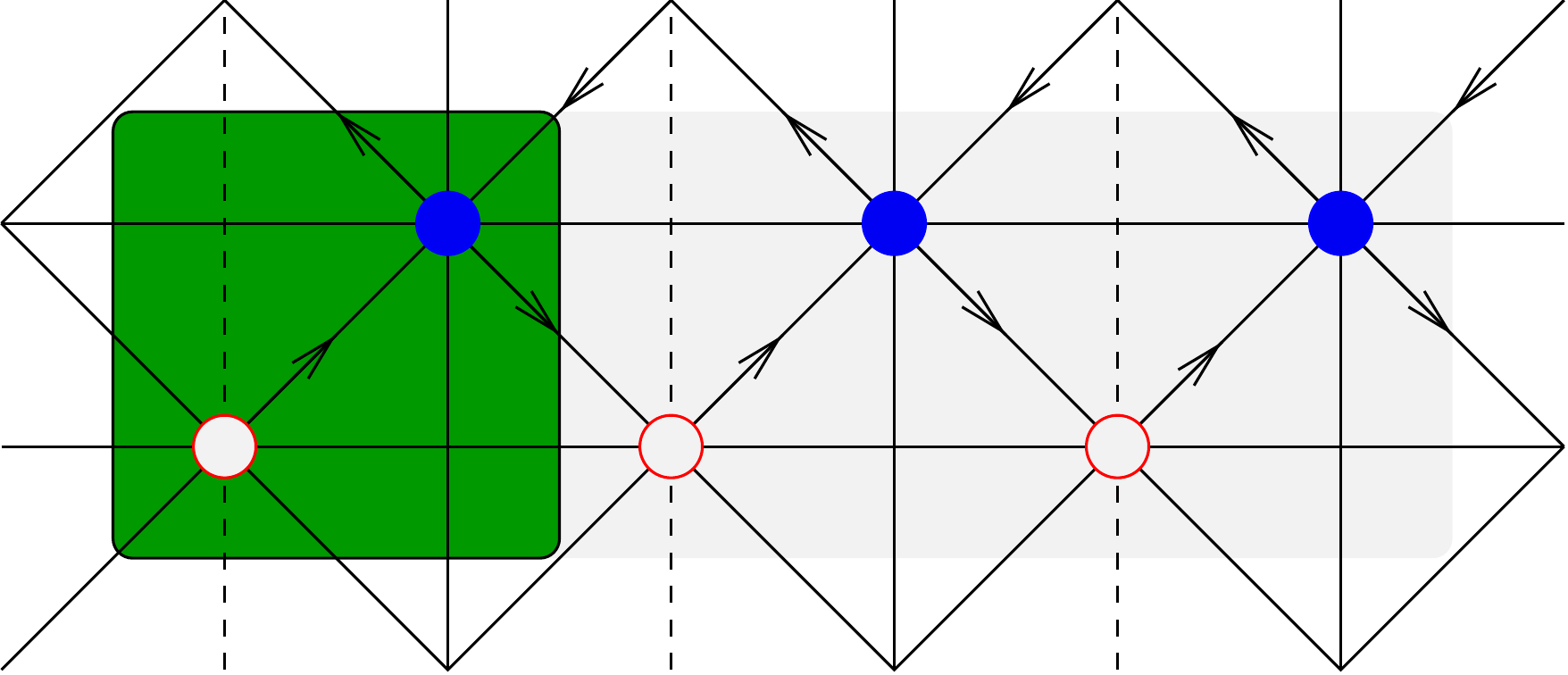}
\end{center}
\caption{
\label{fig:cuberoot}
The checkerboard flat band model and its cube root.
The original unit cell is shaded in green;
the $3\times$-enlarged unit cell is less shaded.
The arrows indicate the direction in which the hopping amplitude is $e^{i \varphi}$.
$t_1'$ ($t_2'$) is the NNN amplitude associated with the solid (dashed) lines.}
\end{figure}

The structure of the lattice is
indicated in Fig.~\ref{fig:cuberoot} (see Fig. \ref{fig:checkerboard_3d} for a different view).  We now further specialize our notation to the checkerboard case by explicitly distinguishing the two sublattices in $\mathcal{V}$.
More explicitly, define Fourier modes for the $a$ and $b$ sublattices (empty
and filled circles in Fig. \ref{fig:cuberoot}):
\be
  a_k = \frac{1}{\sqrt{N_a}} \sum_{x_a} e^{ik x_a} c_{x_a} , ~~~
  b_k = \frac{1}{\sqrt{N_b}} \sum_{x_b} e^{ik x_b} c_{x_b} ~~,
\ee
where $N_a$ and $N_b$ are the number of sites in the $a$ and $b$ sublattices.
The Hamiltonian is $H = \sum_{k \in BZ}  \( h_{2N} + h_{3N} + h_{4N}\)$ with
\begin{align}
h_{2N} &= - t e^{ i \varphi} \left[ b^\dagger_k a_k \( e^{ {i \over 2} \( k_x + k_y\) }
+  e^{ {i \over 2} \( - k_x + k_y\) } \)   \right. &\\
& \left.   a^\dagger_k b_k
\( e^{ {i \over 2} \( k_x - k_y\) } + e^{ {i \over 2} \( -k_x + k_y\) }
\)  \right] + {\text{h.c.}}   &
\nonumber
\\
\nonumber
h_{3N} &= - a^\dagger_k a_k \(t_1'   e^{i k_x} + t_2' e^{i k_y} \)
- b^\dagger_k b_k \( t_2'  e^{i k_x} + t_1' e^{i k_y} \)
&\\
\nonumber
h_{4N} &= - t'' \( a^\dagger_k a_k + b^\dagger_k b_k \) \( e^{ i (k_x+ k_y) } + e^{ i (-k_x + k_y)} \)
&.
\end{align}

Because the models we consider have fractional filling for the microscopic bosons and electrons, we must enlarge the unit cell of the partons to achieve a gap in the single particle parton spectrum.  However, we wish to insure that the original translation symmetry for gauge invariant variables is preserved (see Ref.~\cite{LuRan} for a thorough description of this requirement and its relation to the projective symmetry group).  As an aside, we note that this is not a fundamental requirement since explicit translation symmetry breaking and fractionalization can coexist, but because of its simplicity and relevance to recent numerical efforts, we choose to isolate the fractionalization physics.  Thus we assign the discrete phases $\omega_{rr'}$ in a way that doubles (bosons, $n=2$) or triples (electrons, $n=3$) the unit cell.  We will label the subcells within the enlarged unit cell using script letters $\aa,\bb,...$. Our rule for assigning the phases $\omega_{rr'}$ is as follows: when leaving the $\aa$th subcell of the enlarged unit cell, particles acquire a phase $\omega_n^\aa$, where $\omega_n \equiv e^{2 \pi i/n}$. This has the interpretation as the spontaneous development of a magnetic field of the slave gauge field, specifically, a field in the center of the gauge group.  More explicitly, define clock matrices in the
enlarged-unit-cell basis $\aa, \bb = 1..n$:
\be\label{eq:clock}
\Omega(\alpha)_{\aa \bb} \equiv
e^{i \alpha {2\pi \aa \over n} } \delta_{\aa \bb}
.
\ee

We first rewrite the original Hamiltonian in the new basis, so far a purely cosmetic transformation.
The original Hamiltonian is now a sum over the reduced Brillouin zone $BZ'$:
\be\label{eqn:defofhh}
H = \sum_{k \in BZ'} \( a^\dagger, b^\dagger\)_\aa \hh_{\aa\bb}(k) \(a,b\)^T_{\bb}~.
\ee
Now we take the $n$th root.  Our definition of the $n$th root tightbinding model consists of replacing $\hh$ in
\eqref{eqn:defofhh} by
\be
\hh_{n\text{th root}} =  \hh_{2N}\cdot \Omega(\alpha)
+ \hh_{3N}\cdot \Omega(\alpha')
+ \hh_{4N}\cdot \Omega(\alpha'') ~;
\ee
the parameters $\vec \alpha = (\alpha, \alpha', \alpha'')$
run from $1$ to $n$ and
represent independent choices of $n$th root for each type of hopping.
The resulting bandstructure for
$n=2$ and one choice of $\alpha$s is shown in
Fig.~\ref{fig:sqrt-checkerboard-bandstructure}.

As we have repeatedly stated, the trick is to have the quadratic parton Hamiltonian break the translation group up to gauge transformations.  The simplest such Hamiltonians, realized in our examples below, are proportional the identity on the color indices.  We may make use of discrete fluxes to obtain gapped topological bands, but these lie in the center of $SU(n)$ and don't break the parton gauge group.  A detailed analysis shows that in such cases translation invariance for the electrons requires the same 
flux (mod $2\pi$) through every equivalent loop in the enlarged unit cell.  A simple way to see this is to note that the electron interactions, as generated by the strong coupling expansion, will not be translation invariant without this constraint.  

More generally, pick a parton Hamiltonian $H_1$ that we like, which gives nice gapped topological bands, but whose fluxes do not preserve the original unit cell.
Define $H_2$ ($H_3$) to be the Hamiltonian obtained by translating all the hoppings by one (two) original unit cell of the electrons.  Now let the motion of the blue parton be governed by $H_1$, let red move according to $H_2$, and let green move according to $H_3$.  The electron wavefunction is the product of the wavefunctions: $\Psi(z) = sl_1(z) sl_2(z) sl_3(z)$.  By construction, translation by one original unit cell $T$ acts by $T sl_1 = sl_2, T sl_2 = sl_3, T sl_3 = sl_1$ and therefore $T \Psi = \Psi$.

More formally, our new hamiltonian can be described as follows. Our old hamiltonian was $H_{old} = \sum_{\alpha = 1..3} h_1 c_\alpha^\dagger c_\alpha$.
where $\alpha$ is the color index.  The new Hamiltonian is $H_{new} = \sum_{\alpha} h_\alpha c_\alpha^\dagger c_\alpha$.  The only catch is that our new Hamiltonian no longer preserves the whole $SU(3)$ parton gauge group.  The unbroken gauge group is
$$ G = \{ U \in SU(3)\, \mbox{ s.t. } H_{new} \mbox{ is preserved under } c_\alpha \to U_\alpha^\beta c_\beta \}.
$$
Using this definition, the subgroup of $SU(3)$ that preserves the new Hamiltonian is
$U(1) \times U(1)$.  
This group acts by various diagonal phase rotations of the parton colors.
The action of translation by one original-unit-cell, accompanied 
by a permutation of the colors (the Weyl group of $SU(3)$)
is a projective global symmetry.
The resulting Chern-Simons theory 
is $U(1)_4 \times U(1)_6$,
which has the same topological properties as the full $SU(3)$ theory.

\subsection{Bosons at $1/2$ filling}

 \begin{figure}[h] \begin{center}
  \includegraphics[height=90pt]{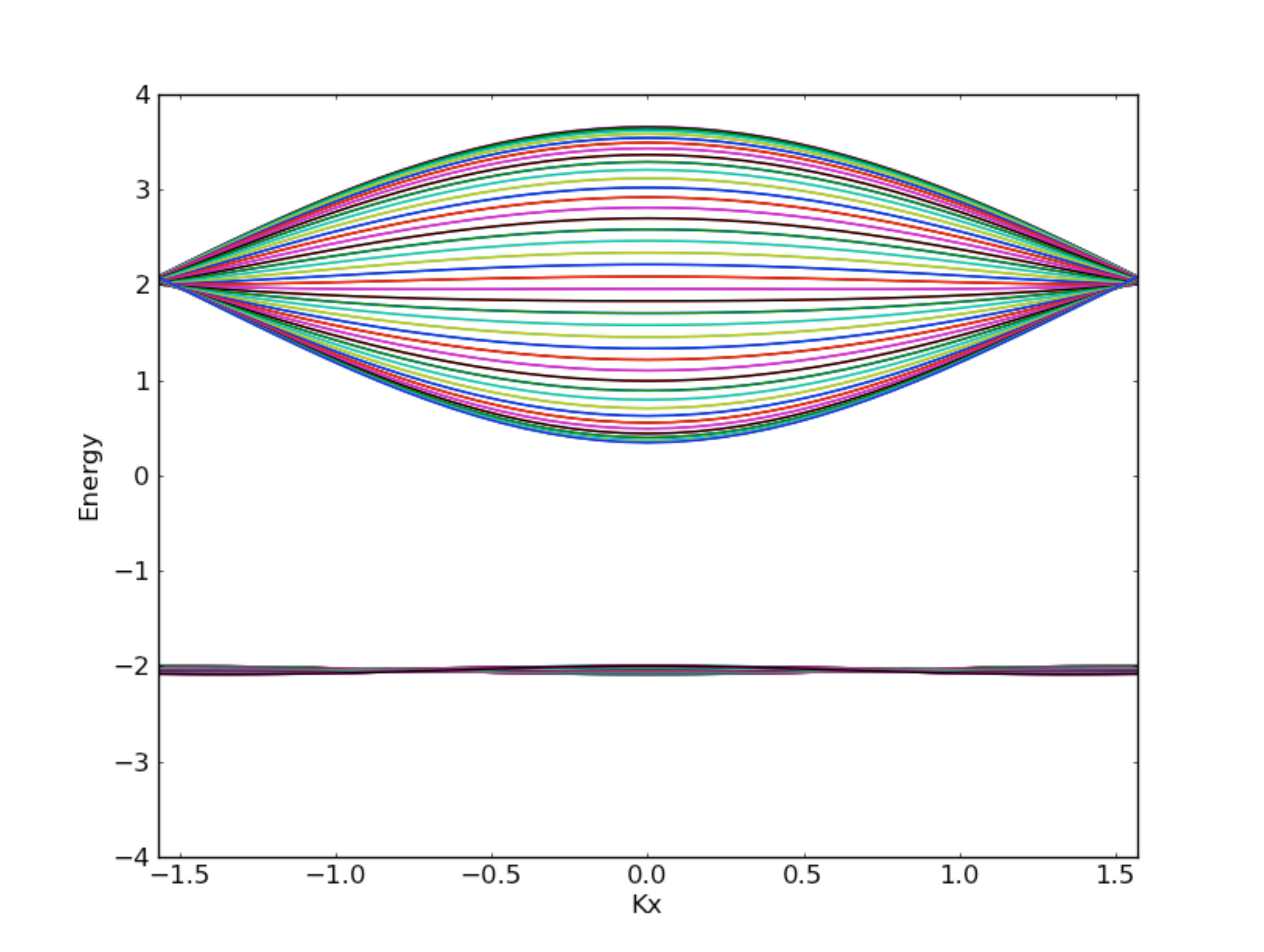}
\includegraphics[height=90pt]{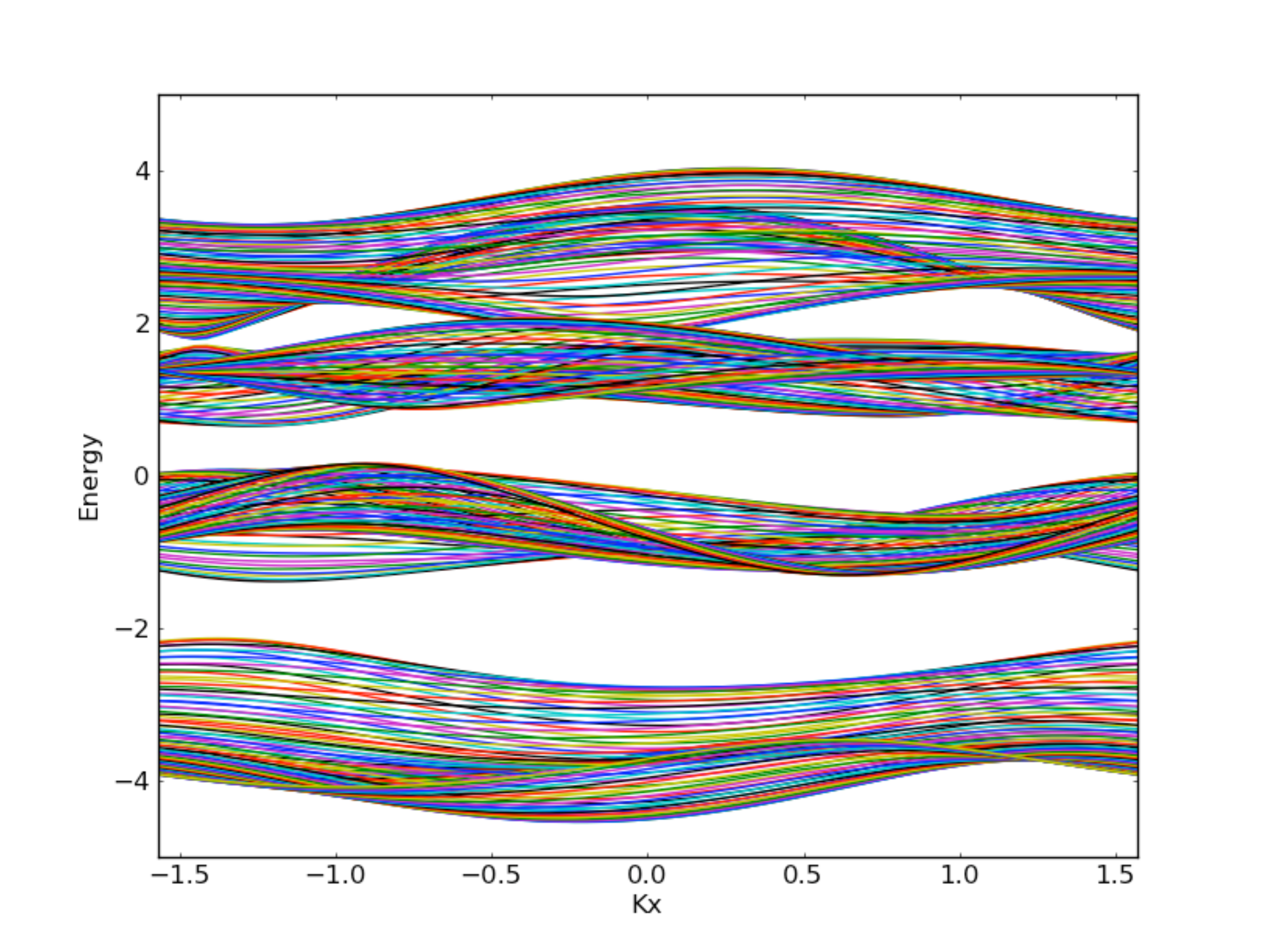}
\end{center}
\caption{
\label{fig:sqrt-checkerboard-bandstructure}
Left: The topological flat bands found in the checkerboard model by \cite{Sun10},
at the optimal hopping amplitudes.  Energy is measured in units of the NN amplitude $t$,
and wavevectors are measured in units of the inverse NNN lattice spacing.  The Hamiltonian is the one used
in the numerical work \cite{Neupert, Sheng11, evelyn, hardcore-bosons, bernevig},
which is {\it minus} that studied in \cite{Sun10}.
Right: The bandstructure for a square root of the checkerboard flat band model.
Each parton hopping amplitude is the square root of the optimal-flatness values
chosen in \cite{Sun10}:
$
 t = 1,~
 t_1' = \sqrt{1 \over 2 + \sqrt{2}}, ~
 t_2' = t_1' e^{i \pi \over 2}, ~
 t'' = i\sqrt{1 \over 2+ 2 \sqrt{2}}, ~
 \varphi = 5 \pi/8,
 \vec{\alpha} = (1,1,0), \vec\theta=(0,1,1,1);
$.
The discrete parameters $\vec \theta$ is explained in appendix B.
The Chern number of the bottom band is $-1$.}
\end{figure}

Since the hardcore bosons are at $1/2$ filling,
and the boson creation operator is $b^\dagger = f_1^\dagger f_2^\dagger$,
each color of parton is also at $1/2$ filling.
Following the procedure outlined above,
we find bandstructures like the one in Fig.~\ref{fig:sqrt-checkerboard-bandstructure}
where the lowest band is separated by an energy gap
from the other four. The mean-field groundstate for each parton species is then a Slater determinant
of the states in the lowest band. The candidate boson groundstate wavefunction is the projection of this state
onto the gauge singlet sector, a procedure motivated by the strong coupling limit in the gauge theory. In this case, where the whole SU$(2)$ gauge symmetry is preserved by the mean field parton Hamiltonian, the boson wavefunction is just the square of the Slater determinant.  We see that it naturally incorporates both the physics of the Chern band, since the partons have a Chern number, as well as the physics of strong correlation, since the boson wavefunction, as the square of a fermion wavefunction, forbids the bosons from approaching each other.

The sum of the Chern numbers of the filled bands for each parton color is $1$.
In the continuum limit they therefore produce a Chern-Simons term
for the SU$(2)$ gauge field with coefficient $k=1$.
The effective field theory is SU$(2)$ level 1 Chern-Simons gauge theory,
which is related by level-rank duality to $U(1)$ level $2$.
This state exhibits charge-1/2 quasiparticles
with anyonic statistics.  Furthermore, each parton color has a Hall conductivity of $\(\frac{e}{2}\)^2 \frac{1}{h}$, and with two colors, we have a total Hall conductivity of $\frac{1}{2} \frac{e^2}{h}$.  More generally, an $n$ color parton model will have a Hall conductivity of $\frac{1}{n} \frac{e^2}{h}$ for the Abelian phases considered here.

In the bandstructure shown in Fig.~\ref{fig:sqrt-checkerboard-bandstructure} we have followed very literally the leading-order result of the
strong coupling expansion.  If we permit ourselves to treat the absolute values of the parton hopping ampitudes as variational parameters (the phases are sacred), we can achieve flatter parton bands with larger bandgaps. This freedom should be kept in mind in future variational studies using these parton wavefunctions.  We note that this bandgap matches onto the quasiparticle gap in the low-energy theory.

\subsection{Fermions at $1/3$ filling}
\begin{figure}[h] \begin{center}
\includegraphics[height=110pt]{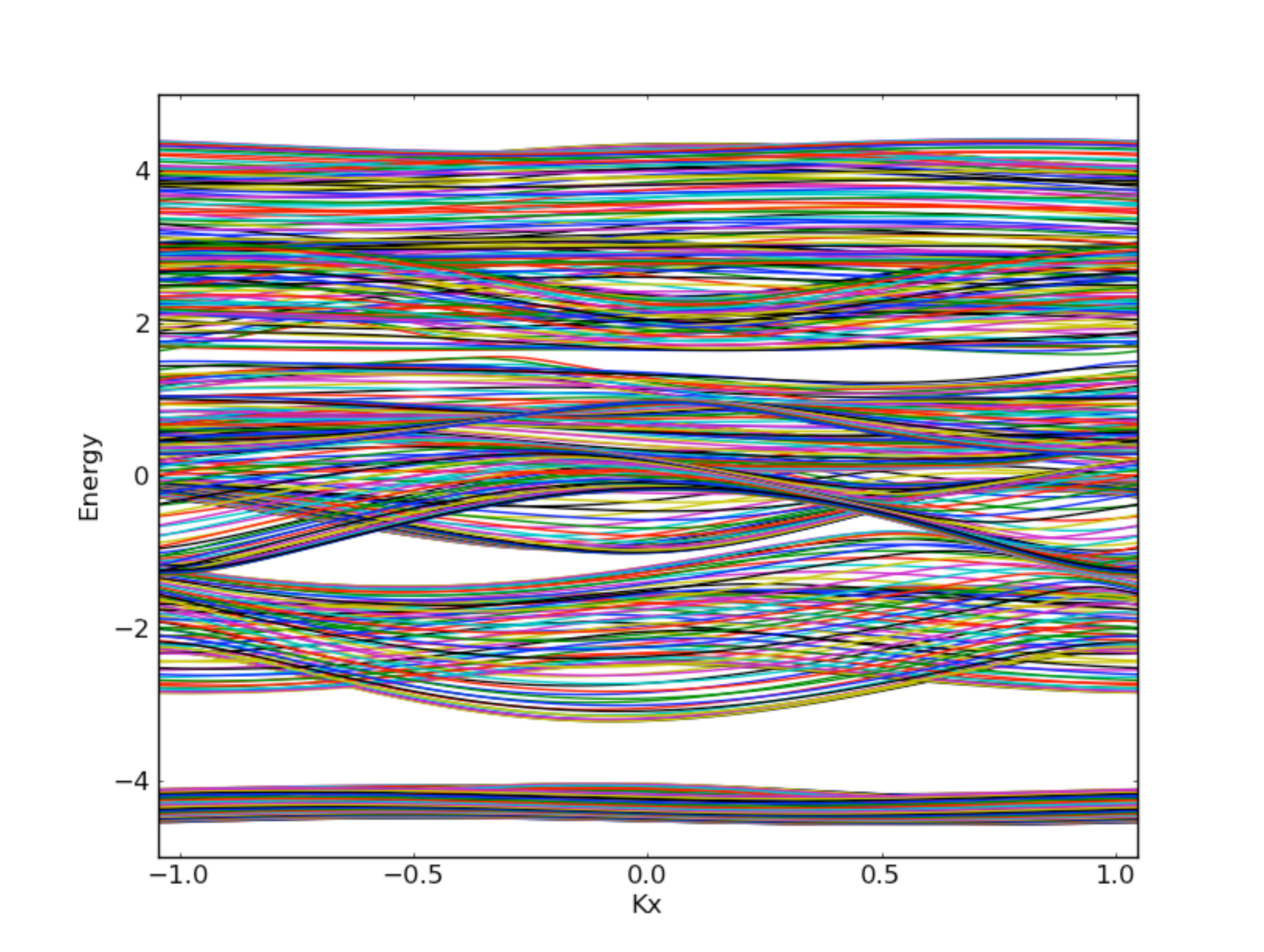}
\end{center}
\caption{
\label{fig:cuberoot-checkerboard-bandstructure}
The bandstructure for a cube root of the checkerboard flat band model.
Each parton hopping amplitude is the square root of the optimal values
chosen in \cite{Sun10}:
$
 t = 1,~
 t_1' = \sqrt[3]{1 \over 2 + \sqrt{2}}, ~
 t_2' = t_1' e^{- {\pi i \over 3}}, ~
 t'' = \sqrt[3]{1 \over 2+ 2 \sqrt{2}}, ~
 \varphi = \pi/12,
 \vec{\alpha} = (1,1,0)  ~
$.
The Chern numbers of these bands, from bottom to top,
are $(1, 0, -1, 0, 0, 0)$.
}
\end{figure}

A very similar story obtains for the checkerboard flatband model
filled with a (spinless) electron for every six sites.
Following the procedure detailed in the previous section with $n = 3$,
we construct a family of cube roots of the checkerboard tightbinding model,
parameterized by the discrete phases $e^{ 2\pi i \vec \alpha/n}$.
A favorable result is shown in Fig.~\ref{fig:cuberoot-checkerboard-bandstructure}.
The cube of the Slater determinant of the lowest band is our candidate wavefunction.  As before, the wavefunction incorporates both strong correlation and Chern band effects.  It is precisely analogous to Laughlin's model wavefunction for $\nu = 1/3$ (which is also the cube of a Slater determinant).

\section{Conclusions}
We have provided model wavefunctions for fractional Chern insulators using the parton approach.  We also analyzed the parton gauge theory in a strong coupling expansion and obtained a mapping to a microscopic electron model.  Furthermore, we argued that both sides of this mapping are under control, based on a combination of numerical, experimental (in equivalent fractional quantum Hall systems), and analytical results.  Thus we have a rather special situation where the parton approach to fractionalized phases can reliably be used, even at the lattice scale.

We find it amusing to note that the construction we have described is a cognate of various mechanisms for fractionalizations
of momentum \cite{MaldacenaSusskind} and D-brane charge quantum numbers \cite{DouglasMoore} in string theory. Like the topology of bandstructure,
the latter enjoys a K-theory classification \cite{Witten}.

In the future, we believe it would be very interesting to study the gauge theory numerically, as well as to compare our wavefunctions to those produced in the recent numerical calculations.  It would also be interesting to find a Hamiltonian for which our state is the exact ground state, either by pushing the strong coupling expansion or by using an analog of Haldane's pseudopotentials.  We are currently pursuing lattice models that might realize non-Abelian fractional Chern insulators, and we anticipate that our strong coupling expansion will be useful in this endeavor.  Finally, it would be quite interesting to propose analogous realistic models in 3d that might realize fractional topological insulators.  This can be achieved as shown in Ref.~\cite{PhysRevB.83.195139} by spontaneously breaking the gauge symmetry in the mean field parton Hamiltonian.  We can also consider models in $2+1$ dimensions that break the SU$(n)$ gauge group down to some subgroup, {\it e.g.}~a combination of U$(1)$s or a discrete group such as ${\bf Z}_n$.  These models are realized at the mean field level by parton Hamiltonians in which the different colors experience different hoppings, and in the low energy limit, they may be described by Chern-Simons-Higgs theories.  However, as we indicated above, some care must be taken with lattice symmetries in such broken phases.

More generally, our lattice gauge theory and strong coupling expansion provide powerful tools to address outstanding questions.  We give three examples.  First, it is experimental lore that in realistic materials the gap separating a flat electron band will be on the order of the spin-orbit coupling.  Since interactions are likely to be at the Coulomb scale, it is potentially unlikely that the interaction energy is less than the electronic band gap.  This brings in the danger of mixing with levels with different Chern number, potentially complicating the physics.  From the perspective of our gauge theory, this situation can be modeled by adding bare baryon-baryon interactions. So long as these interactions are small compared to the bare parton gap $\sim t^f$, they do not appear to cause a phase transition, but this criterion is compatible with interactions that are larger than the bare electron gap.  The story is only slightly complicated by the fact that the gap to quasiparticles in the gauge theory should not be set by $t^f$ but by some much smaller renormalized energy scale (one commensurate with electronic energy scales). Second, our mapping from the parton gauge theory to the electron model also suggests that we could have models where the bare electron band does not have a Chern number (despite breaking time reversal), and yet the partons remain in gapped Chern bands.  Third, our mapping naturally provides a way to take a parton band structure that realizes a Chern number two band and guess a microscopic electron model.  This case is interesting because an $SU(3)$ level 2 topological phase is non-Abelian and universal for quantum computation.

Since it is of particular interest, we comment a bit more on our progress towards non-Abelian states.  Regarding the checkerboard model, while individual bands occasionally have Chern numbers with absolute value larger than $1$, we have not found choices of $\vec \alpha$ (or the other order-$n$ ambiguities in the $n$th root procedure) that lead to bandstructures where the sum of the Chern numbers of the lowest bands adds up to a number with absolute value larger than $1$. However, the model studied by Hatsugai and Kohmoto \cite{HatsugaiKohmoto, Oshikawa94} can be viewed as a cube root of the square lattice with NNN hoppings which spontaneously breaks time reversal symmetry. In a range of parameters, the lowest band has Chern number $2$.  This suggests that (spinless) electrons on this square lattice at 1/3 filling and repulsive interactions will form a non-Abelian FQH state.  We are in the process of investigating this claim further.

\vskip.2in
{\bf Acknowledgements}
We thank M.~Barkeshli, E.~Tang, D.~Tong and X.~G.~Wen for discussions and encouragement.
The work of JM is supported in part by
funds provided by the U.S. Department of Energy
(D.O.E.) under cooperative research agreement DE-FG0205ER41360,
and in part by the Alfred P. Sloan Foundation.  BGS is supported by a Simons Fellowship through Harvard University.

\appendix
\section{Appendix A: Perfectly flat Chern bands}
It is possible to have a perfectly flat band with non-zero Chern number provided we relax the assumption of finite range interactions.  Consider for simplicity a gapped hamiltonian $H$ with two bands with opposite and non-zero Chern numbers.  Our construction will also work for more complicated Hamiltonians.  Let $P$ be the projector onto the lowest band.  The hamiltonian $H_{\mbox{flat}} = H - P H P$ has a perfectly flat lower band with the same Chern number as before.  The question we must answer is: how non-local is the operator $P$?

Consider a function $f(t)$ whose fourier transform $\tilde{f}(\omega)$ has the property that $\tilde{f} = 1$ for $\omega$ in the lower band and $\tilde{f} = 0$ for $\omega $ in the upper band.  $f$ is otherwise arbitrary, although we will want it to decay as fast as possible at large $t$.  Since we require that the Fourier transform of $f$ vanish outside a compact domain, $f$ cannot decay exponentially fast in $t$ (otherwise the Fourier transform would be analytic in a strip).  However, it can decay faster than any polynomial and nearly exponentially.

Now we form the operator
\be \tilde{P} = \int_{-\infty}^{\infty} dt f(t) e^{ i H t} = \sum_{E_n} \tilde{f}(E_n) |E_n \rangle \langle E_n |
.\ee  However, this operator is nothing but the projector $P$ because of the properties of $\tilde{f}$.  We are now in a position to evaluate $\langle r | P | r' \rangle$ where $r r'$ are positions on the lattice.  The key point is that although we integrate over all times, long times receive an extremely small weighting.  At early times, the amplitude to go from $r'$ to $r$ is very small because the particle hasn't had a chance to move.  At late times, the weighting factor $f$ is very small.  Thus there is some intermediate time that dominates the matrix element, and the matrix element can be quite small if the separation between $r$ and $r'$ is large.  The characteristic timescale in $f$ is the inverse gap $\Delta^{-1}$.  Assume for simplicity that the propagation under $H$ is ballistic with speed $v$.  Then the transition amplitude is very small for $|r-r'| \gg v t$ and the weight factor is very small if $\Delta t \gg 1$.  Combining these two facts, we expect a sharp decay in the matrix element of $P$ once $|r-r'| \gg v/\Delta$.  This decay will be almost exponential as described above.

Assembling everything together, the operator $P H P$ only delocalizes the terms in $H$ by a size of roughly $v/\Delta$ up to corrections that are almost exponentially small.  The hopping terms in $P H P$ will not be of strictly finite range, but they will decay very rapidly.
For more information about these techniques, see \cite{hastings}.
Of course, we have not proved that there is no finite range model with a perfectly flat Chern band.

Regarding the gauge theory on a lattice with such long range hoppings, we are free to abandon our original procedure of associating a new $SU(n)$ link variable with every hopping.  Instead, we may simply stretch a Wilson line built of existing gauge variables to connect the distant sites.  This only modifies our strong coupling calculation by a factor of $1/x$ where $x$ is the lattice length of the Wilson line.  However, this modification may be reincorporated into the long range hoppings without affecting their super-polynomial decay.  Gauge kinetic terms involving these long Wilson lines will be generated upon integrating out the fermions, but we may expect them to have a suitably small prefactor.  We can find no serious conceptual issues with the gauge theory defined with such long range hoppings, but we are aware that we cannot rule out subtle pathologies.  The fact that the corresponding electron model appears to reliably enter a fractionalized phase described by a conventional low energy gauge theory suggests that the inclusion of these long range hoppings is not a serious modification of the physics.
\section{Appendix B: Explicit $n$th root Hamiltonians}

Here we give more explicit expressions for the parton hopping matrices
in the checkerboard lattice models.
In the following, the parton gauge indices are omitted because the Hamiltonians
we describe are proportional to the identity matrix in that space.
It will be convenient to define $k_1 \equiv k_x+ k_y$ and $k_2 \equiv -k_x + k_y$.
In each case, the tightbinding Hamiltonian is a sum of nearest-neighbor (NN), next-nearest-neighbor (3N)
and next-next-nearest-neighbor (4N) hopping terms.
For reference, the original checkerboard model
\cite{Sun10}
is
\be
H = -\sum_{BZ}(a^\dagger, b^\dagger) \( \hh_{2N} + \hh_{3N} + \hh_{4N} \)
(a, b)^T + \text{h.c.} ,
\ee
where
\be
 \hh_{2N} = t e^{i \phi}  \left( \begin{array}{cc}
0 & e^{i{k_2 \over 2}}  + e^{ - i {k_2 \over 2} } \\
e^{i {k_1 \over 2}} + e^{ - i {k_2 \over 2}}& 0
 \end{array} \right)\ee
\be
\hh_{3N} = e^{i k_y} \left( \begin{array}{cc}
t_2' & 0 \\
0 & t_1'  \end{array} \right) +
 e^{i k_x}\left( \begin{array}{cc}
t_1' & 0\\
0 & t_2'\end{array} \right) \ee
\be
 \hh_{4N} = t'' (e^{i k_1} + e^{i k_2}) \left( \begin{array}{cc}
1 & 0 \\
0 & 1 \end{array} \right).
\ee

\subsection{Square Root of the (upside down) Checkboard model}

\bwt
To begin, define building-block matrices in the basis $(a_{1}, b_{1}, a_{2}, b_{2})$,
where $\aa=1,2$ here is the subcell index:
\be
 \hh_{2N} = t \left( \begin{array}{cccc} 
0 & e^{i({k_2 \over 2}+\phi)} & 0 & 0\\
e^{i (\phi+{k_1 \over 2})} & 0 & 0 & 0  \\
0 & e^{i(-{k_2 \over 2}+\phi + \pi \theta_1)}+e^{i(-\phi+{k_1 \over 2}+\pi \theta_2)} &  0 & e^{i({k_2 \over 2}+\phi)} \\
e^{i(\phi-{k_1 \over 2} +\pi \theta_3)}+e^{i (-\phi+{k_2 \over 2}+\pi \theta_4)} & 0  & e^{i({k_1 \over 2}+\phi)} & 0 \end{array} \right)\ee
\be
\hh_{3N} = e^{i k_y} \left( \begin{array}{cccc}
t_2' & 0 & 0 & 0\\
0 & t_1' & 0 & 0 \\
0 & 0 &  t_2 ' & 0 \\
0 & 0 & 0 & t_1 ' \end{array} \right) +
 e^{i k_x}\left( \begin{array}{cccc}
0 & 0 & t_1' & 0\\
0 & 0 & 0 & t_2' \\
t_1' & 0 &  0 & 0 \\
0 & t_2' & 0 & 0 \end{array} \right) \ee
\be
 \hh_{4N} = t'' (e^{i k_1} + e^{i k_2}) \left( \begin{array}{cccc}
0 & 0 & 1 & 0 \\
0 & 0 & 0 & 1 \\
1 & 0 &  0 & 0 \\
0  & 1 & 0 & 0 \end{array} \right)
\ee
For $n=2$, the clock matrices discussed in Eqn.~\eqref{eq:clock}
take the explicit form
\be \Omega(\alpha) = \left( \begin{array}{cccc}
1 & 0 & 0 & 0 \\
0 & 1 & 0 & 0 \\
0 & 0 &  e^{i \alpha \pi} & 0 \\
0  & 0 & 0 & e^{i \alpha \pi} \end{array} \right).
\ee
Then, the general Hamiltonian for the square-root of the checkerboard model is
\be
H = -\sum_{BZ}(a_1^\dagger, b_1^\dagger, a_2^\dagger, b_2^\dagger)(\vec{\hh} \cdot \vec{\Omega})
(a_1, b_1, a_2, b_2)^T + \text{h.c.} ,
~~~~\text{with}~~\vec{h} \equiv (h_{2N}, h_{3N}, h_{4N}), ~~
\vec{\Omega} \equiv (\Omega(\alpha), \Omega(\alpha'), \Omega(\alpha '')).
\ee

The parameters $\alpha, \alpha', \alpha''$ and $\theta_{1,2,3,4}$ take values $1..n=2$ and parametrize
the ambiguities in taking the square root of each hopping amplitude.
Note that
we could introduce analogs of the $\theta$ parameters for the 3N and 4N hoppings as well;
we have not yet explored this possibility.

The parameters used in Fig.~\ref{fig:sqrt-checkerboard-bandstructure} are
\be
 t = 1, ~~
 t_1' = \sqrt{1 \over 2 + \sqrt{2}},~~
 t_2' = t_1' e^{i \pi \over 2}, ~~
t'' = i\sqrt{1 \over 2+ 2 \sqrt{2}}, ~~
 \phi = {5 \pi \over 8},~~
 \vec{\alpha} = (1,1,0), ~~
 \vec{\theta} = (0,1,1,1)~.
\ee

\subsection{Cube Root of the Checkerboard Model}

Similarly, the cube root tightbinding model is constructed as follows.
\be
\hh_{2N} =  t e^{i \phi} \left( \begin{array}{cccccc}
0 & e^{ik_2 \over 2} & 0 & 0 & 0  & e^{-i k_2 \over 2} \\
e^{i k_1 \over 2} & 0 & e^{- i k_1 \over 2} & 0 & 0 & 0 \\
0 & e^{-i k_2 \over 2} &  0 & e^{ik_2 \over 2} & 0 & 0 \\
0  & 0 & e^{ik_1 \over 2} & 0 &  e^{-ik_1 \over 2} & 0 \\
0 & 0 & 0& e^{-ik_2 \over 2} & 0 & e^{ik_2 \over 2} \\
e^{-ik_1 \over 2} & 0 & 0 & 0 & e^{ik_1 \over 2} & 0\end{array} \right)
\ee
\be
 \hh_{3N} =  e^{ik_y} \left( \begin{array}{cccccc}
t_2' & 0 & 0 & 0 & 0  & 0 \\
0 & t_1' & 0 & 0 & 0 & 0 \\
0 & 0 &  t_2' & 0 & 0 & 0 \\
0  & 0 & 0 & t_1' &  0 & 0 \\
0 & 0 & 0& 0 & t_2' & 0 \\
0 & 0 & 0 & 0 & 0 & t_1' \end{array} \right)
+ e^{ik_x}\left ( \begin{array}{cccccc}
0 & 0 & 0 & 0 & t_1'  & 0 \\
0 & 0 & 0 & 0 & 0 & t_2' \\
t_1' & 0 &  0 & 0 & 0 & 0 \\
0  & t_2' & 0 & 0 &  0 & 0 \\
0 & 0 & t_1' & 0 & 0 & 0 \\
0 & 0 & 0 & t_2' & 0 & 0 \end{array} \right)
\ee
\be \hh_{4N} =  t''(e^{ik_1}+ e^{-ik_2}) \left( \begin{array}{cccccc}
0 & 0 & 0 & 0 & 1  & 0 \\
0 & 0 & 0 & 0 & 0 & 1 \\
1 & 0 &  0 & 0 & 0 & 0 \\
0  & 1 & 0 & 0 &  0 & 0 \\
0 & 0 & 1& 0 & 0 & 0 \\
0 & 0 & 0 & 1 & 0 & 0 \end{array} \right)
\ee
We introduce the $\Omega$ matrices similarly:
\be \Omega(\alpha, \beta)= \left( \begin{array}{cccccc}
1 & 0 & 0 & 0 & 0  & 0 \\
0 & 1 & 0 & 0 & 0 & 0 \\
0 & 0 &  e^{i 2 \pi \alpha \over 3} & 0 & 0 & 0 \\
0  & 0 & 0 & e^{i 2 \pi \alpha \over 3} &  0 & 0 \\
0 & 0 & 0 & 0 & e^{i 2 \pi \beta \over 3}& 0 \\
0 & 0 & 0 & 0 & 0 & e^{i 2 \pi \beta \over 3} \end{array} \right) .\ee
\be\label{eq:cuberoot}
H = + \sum_{BZ}(a_1^\dagger, b_1^\dagger, a_2^\dagger, b_2^\dagger, a_3^\dagger, b_3^\dagger)(\vec{\hh} \cdot \vec{\Omega})
(a_1, b_1, a_2, b_2, a_3, b_3)^T + {\text{h.c.}},
\ee
with $ \vec \hh$ and $\vec \Omega$ defined as above.
This notation is slightly more general than that used in the body of the paper;
the set of parameters we used for the cube root in Fig.~\ref{fig:cuberoot-checkerboard-bandstructure}
translates to
\be
t = 1,~~
t_1' = \sqrt[3]{1 \over 2 + \sqrt{2}},~~
t_2' = t_1' e^{-i \pi \over 3},~~
t'' = \sqrt[3]{1 \over 2+ 2 \sqrt{2}},~~
\phi = {\pi \over 12},~~
\vec{\alpha} = (1,1,0),~~
\vec{\beta} = (2,2,0)~.
\ee
where $ \vec{\alpha} \equiv (\alpha, \alpha', \alpha ''),
 \vec{\beta} \equiv (\beta, \beta', \beta'')$.
Note that the overall sign in \eqref{eq:cuberoot}
accounts for the sign-reversal relative to \cite{Sun10}
which puts the flat band on the bottom,
using the key equation $(-1)^3=-1$.

\ewt

\bibliography{fci_wf}
\end{document}